\documentclass[9pt,twocolumn,twoside]{osajnl}
\usepackage{url}
\usepackage{breakurl}

\usepackage{float}
\usepackage{textcomp}
\usepackage{upgreek}

\journal{ao} 

\setboolean{shortarticle}{true}

\title{Optical frequency generation using fiber Bragg grating filters for applications in portable quantum sensing}

\author[1,*]{Calum D. Macrae}
\author[1]{Kai Bongs}
\author[1]{Michael Holynski}

\affil[]{University of Birmingham, School of Physics and Astronomy, Edgbaston, Birmingham, B15 2TT}

\affil[*]{Corresponding author:calumdonaldmacrae@gmail.com}

 \doi{\url{}}

\begin{abstract}
A method for the agile generation of the optical frequencies required for laser cooling and atom interferometry of rubidium is demonstrated. It relies on fiber Bragg grating technology to filter the output of an electro-optic modulator and was demonstrated in a robust, alignment-free, single-seed, frequency-doubled, telecom fiber laser system. The system was capable of 50 ns frequency switching over a \char`~40 GHz range, \char`~0.5 W output power and amplitude modulation with a \char`~15 ns rise/fall time and an extinction ratio of 120 $\pm$ 2 dB.  The technology is ideal for enabling high-bandwidth, mobile industrial and space applications of quantum technologies.  
\end{abstract}

\setboolean{displaycopyright}{false}

\usepackage{url}

\begin{document}

\maketitle

\section{Introduction}
\label{sec:intro}
Since its inception in 1991\cite{atomint}, atom interferometry has been used to measure rotation\cite{rotation}, gravitational acceleration\cite{gravity}, gravity gradients\cite{gradio} and to test the Equivalence Principle\cite{equiv}. Given sensing based on atom interferometry has demonstrated exceptionally low instrument drift and suppression of vibrational noise\cite{mobilegrav}, there is particular interest in developing portable quantum devices. These could enable future applications of atom interferometry in civil engineering\cite{civil} and space-based sensors\cite{space2}. 

Generally, atom interferometry, as well as laser cooling techniques, impose stringent requirements on the laser system. For example: multiple optical frequencies must be generated with mrad phase coherence over the single shot measurement time; optical frequency sweeping is required during laser cooling and to compensate doppler-shift during the freefall of atoms \cite{chirp}; a high amplitude extinction ratio is required to maintain atomic coherence during pulse sequences; and fast optical frequency switching is required to maximize the measurement bandwidth. Furthermore, for portable sensing applications,  compactness and robustness is required. For rubidium atom interferometry, these requirements can be met by using a single-seed, frequency-doubled, telecom fiber laser system.

Several single-seed laser systems, which aim to minimize size and power consumption, have been published for rubidium atom interferometry. In 2015\cite{onboard}, by modulating a 1560 nm seed laser, optical single-sideband modulation over a \char`~1 GHz range in a few ms was achieved. In 2016\cite{compact1560}, an approach was taken where a 1560 nm seed laser was locked to an atomic reference and a Fabry-Perot cavity was used to filter the carrier and undesired sidebands from the output. This allowed optical single-sideband modulation over a \char`~1 GHz range in less than 200 \textmu s. However, only a factor of 5 suppression of the undesired sidebands was achieved, and an optical shutter was used, limiting the bandwidth of pulse generation with a high amplitude extinction ratio. Finally, in 2018\cite{compact780}, a direct 780 nm diode laser approach was used. Although capable of a 578 MHz frequency modulation in 10 ms, the use of tapered amplifiers caused problems in optical power and mode stability. Serrodyne frequency shifting has also been demonstrated as an alternative wideband, single-sideband modulation technique\cite{serrodyne}. However, this requires sawtooth RF waveform generation, which is difficult to implement at high frequencies, resulting in a relatively low suppression of the optical carrier frequency.

When generating multiple optical frequencies, previous single-seed systems used an electro-optic modulator (EOM) to generate optical double-sideband spectra. This serves to decrease the efficiency of optical amplification as power is wasted in the undesired sidebands. This also affects the systems accuracy by introducing additional interactions\cite{extrafreq}. Optical IQ modulation has been demonstrated to avoid such effects\cite{IQ}, but requires relatively complex RF signal generation and drift of the modulator’s bias points must be compensated\cite{IQbias}. 

In this paper, by utilizing fiber Bragg grating (FBG) technology, a method to optical single-sideband modulate over a \char`~40 GHz range with 50 ns frequency jumps is presented. Compared with previous single-seed systems, the method greatly increases the possible Raman detuning from \char`~1 to \char`~30 GHz with a factor of 4000 faster switching.  Unlike previous systems, multiple frequencies can also be generated with a high suppression of all undesired optical frequencies. Furthermore, due to the non-linear efficiency of sum frequency generation\cite{nonlinear}, fast amplitude modulation with a remarkably high extinction ratio can be achieved at 780 nm by modulating at 1560 nm before frequency conversion. This removes the need for optical shutters, greatly increasing the pulse generation bandwidth.

\section{Fiber Bragg Gratings}
\label{sec:FBG}
A fiber Bragg grating is a single mode optical fiber with a periodic modulation of the core refractive index, causing light at the Bragg wavelength to be reflected\cite{braggreview}. As the effective refractive index is dependent on the fiber core temperature and strain\cite{braggreview}, commercial fiber Bragg grating filters are sold in thermally controlled enclosures that stabilize strain and allow thermal tuning of the Bragg wavelength over a \char`~60 GHz range\cite{fbgdata}, while remaining stable to $<$ 1 GHz. They are available with a 50 MHz to 50 GHz reflection bandwidth\cite{fbgdata} which, like the center Bragg wavelength, is fixed at manufacture.  The reflection frequency response is sufficiently sharp to remove sidebands $\geq$ 10 GHz from the output of an EOM with $>$ 20 dB attenuation of the carrier and undesired sidebands at 1560 nm. This is achieved by filtering the output of an EOM using an FBG, as shown in Figure 1. Notably, multiple FBGs can be used to further increase the attenuation of undesired frequencies.
\begin{figure} [H]
\centering
\includegraphics[width=0.85\linewidth]{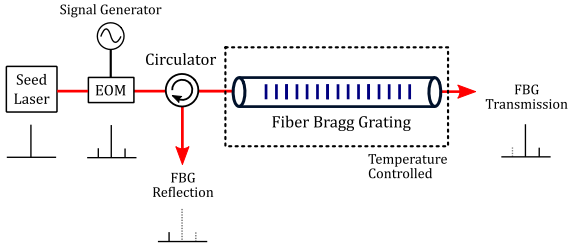}
\caption{Optical filtering of the output of an electro-optic modulator (EOM) using a fiber Bragg grating (FBG). The light reflected from the FBG is separated using an optical circulator. The optical spectrum is depicted at various points in the setup.}
\label{fig:braggfilter}
\end{figure}
\section{Application to atom interferometry}
\label{sec:Application}
By offset locking a seed laser to an atomic transition and tuning a fiber Bragg grating to reflect one set of sidebands from an EOM, single-sideband spectra can be generated over the entire range required for laser cooling and atom interferometry (Fig. 2). This enables agile single-sideband modulation limited only by the electro-optic modulation bandwidth, which can exceed 30 GHz in fiber-coupled EOMs\cite{eomdata}.
\begin{figure} [H]
\centering\includegraphics[width=0.95\linewidth]{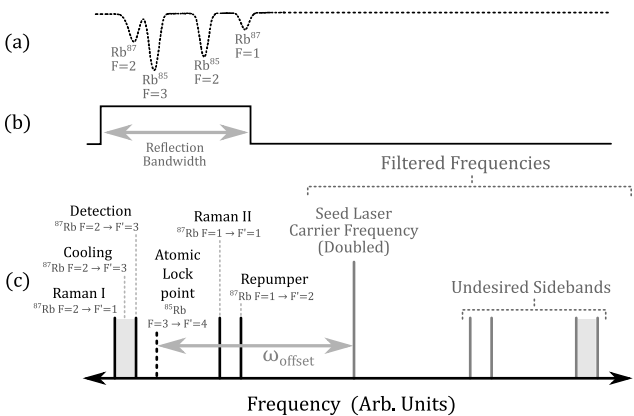}
\caption{The fiber Bragg grating is tuned to reflect only the frequencies required for laser cooling and atom interferometry, and to transmit the carrier and undesired sidebands. (a) The absorption spectrum of the rubidium D\textsubscript{2} line, (b) the ideal reflection response of the fiber Bragg grating and, (c) the frequencies required for laser cooling and atom interferometry. All are shown on the same y-axis.}
\label{fig:filter}
\end{figure}

\subsection{Generating multiple optical frequencies}

Both laser cooling and Raman excitation require two optical frequencies to be generated simultaneously. When these are generated in a single, frequency-doubled laser beam, the process is subject to sum frequency generation. For a laser beam with frequencies, $\Omega_1$ and $\Omega_2$, with electric field amplitudes, E$_1$ and E$_2$, the electric field strength can be described as\cite{nonlinear}:
\begin{equation}
E(t) = E_1 e^{-i\Omega_1 t} + E_2 e^{-i\Omega_2 t} + c.c.
\label{eq:refname1}
\end{equation}
When this is incident on a nonlinear crystal with a second-order electric susceptibility, $\chi^{(2)}$, a nonlinear polarization is created\cite{nonlinear}:
\begin{equation}
P^{(2)}(t) = \epsilon_0\chi^{(2)}E(t)^2
\label{eq:refname2}
\end{equation}
where $\upepsilon_0$ is the permittivity of freespace. This can be shown to generate light at frequencies 2$\upOmega\textsubscript{1}$, $ 2\upOmega \textsubscript{2}$,  $ \upOmega\textsubscript{1}+\upOmega\textsubscript{2}$ and $ \upOmega\textsubscript{1}-\upOmega\textsubscript{2}$\cite{nonlinear}. Assuming only two frequencies, $\upOmega\textsubscript{A}$ and $\upOmega\textsubscript{B}$, are desired at the laser output, and the desired amplitude of $\upOmega\textsubscript{A}>\upOmega\textsubscript{B}$, spectral purity is maximized by setting $\upOmega\textsubscript{A}=2\upOmega\textsubscript{1}$ and  $\upOmega\textsubscript{B}=\upOmega\textsubscript{1}+\upOmega\textsubscript{2}$. Figure 3 shows spectra before and after the nonlinear crystal. 
\begin{figure} [H]
\centering
\includegraphics[width=0.95\linewidth]{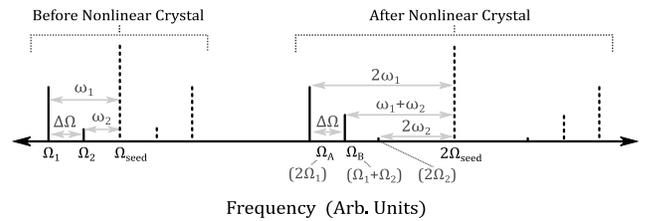}
\caption{Optical spectra before and after the nonlinear crystal. Black and dotted lines denote frequencies reflected and transmitted by the fiber Bragg gratings respectively.}
\label{fig:double}
\end{figure}
For frequency sidebands at RF frequencies, $\upomega\textsubscript{1}$  and $\upomega\textsubscript{2}$, which are applied with an electro-optic modulator at 1560 nm, the optical spectrum after the nonlinear crystal will contain the desired frequencies at $\upOmega\textsubscript{A}= 2\upOmega\textsubscript{seed}-2\upomega\textsubscript{1}$ and $\upOmega\textsubscript{B}=2\upOmega\textsubscript{seed}-(\upomega\textsubscript{1}+\upomega\textsubscript{2})$, where $\upOmega\textsubscript{seed}$ is the optical frequency of the seed laser. Notably, the difference in frequencies, $\Delta\upOmega$, is the same before and after the nonlinear crystal, i.e. $\upOmega\textsubscript{A}-\upOmega\textsubscript{B}=\upOmega\textsubscript{1}-\upOmega\textsubscript{2}$. $\upOmega\textsubscript{seed}$ is locked at $\frac{\upOmega\textsubscript{lock}+\upomega\textsubscript{offset}}{2}$, where $\upOmega\textsubscript{lock}$ is the frequency of the atomic reference and $\upomega\textsubscript{offset}$ is the lock offset frequency.

During laser cooling, the cooling optical frequency is swept whilst the repumper frequency remains fixed. Because the amplitude of the repumper frequency is less than that of the cooling, the cooling and repumper frequencies are generated at $\upOmega\textsubscript{A}$ and $\upOmega\textsubscript{B}$ respectively. Because $\upOmega\textsubscript{B}$  is dependent on both $\upomega\textsubscript{1}$  and $\upomega\textsubscript{2}$, to maintain the repumper at a fixed optical frequency, $\upomega\textsubscript{2}$  must be decreased when $\upomega\textsubscript{1}$  is increased. This was achieved experimentally by generating $\upomega\textsubscript{1}$ and $\upomega\textsubscript{2}$ using RF mixers with a common intermediate frequency, as shown in Figure 4.
\subsection{Implementation in a full laser system}
\label{sec:Implementation}
Figure 4 shows an implementation of the optical filtering scheme in a single-seed laser system used to both laser cool and measure two-photon Raman transitions in ensembles of rubidium atoms.
\begin{figure} [H]
\centering\includegraphics[width=0.95\linewidth]{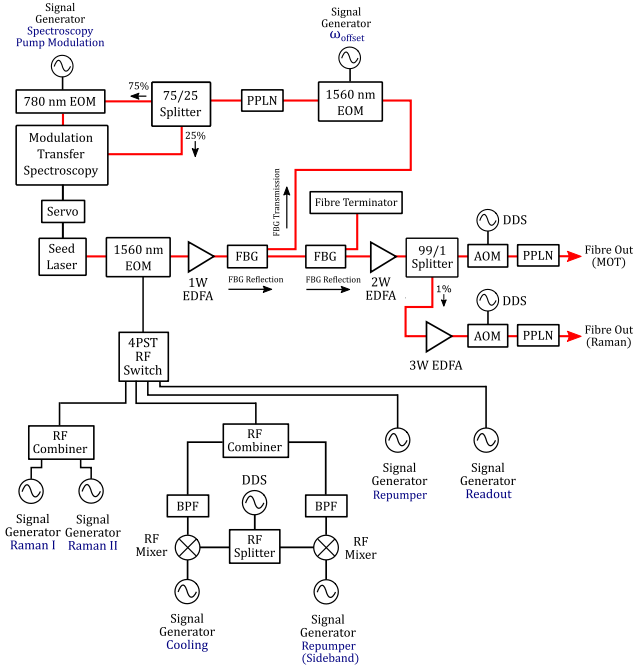}
\caption{Simplified diagram of laser system used to laser cool and measure two-photon Raman transitions in rubidium atom ensembles. The output optical frequency was switched using the quadruple-pole-single-throw (4PST) radio-frequency (RF) switch connected to the electro-optic modulator (EOM). The RF input was filtered with bandpass filters (BPFs) and frequency sweeps were implemented with direct-digital synthesizers (DDS’). The transmission from the first fiber Bragg grating (FBG) was used to offset-lock the seed laser to an atomic reference. The reflection was passed through a second FBG to further attenuate the undesired optical frequencies, then amplified with erbium-doped fiber amplifiers (EDFAs), amplitude modulated with acousto-optic modulators (AOMs) and frequency converted with periodically-poled lithium niobate (PPLN) waveguides.}
\label{fig:laser}
\end{figure}
The transmission from the first FBG (TFN-1560.746-N50-IL3.5-30-C1P-C3, TeraXion), was used to offset frequency lock the seed laser (Koheras Basik E15, NKT) to the F=3 to F’=4 transition of the \textsuperscript{85}Rb D\textsubscript{2} line  ($\upOmega$\textsubscript{lock}) using modulation transfer spectroscopy (MTS)\cite{MTS}. A fiber-coupled MTS setup was built and a servo controller (LB1005-S, Newport) was used to maintain the lock. The atomic offset frequency ($\upomega$\textsubscript{offset})  was \char`~9.1 GHz. Switching between cooling, state preparation, Raman and readout frequencies was achieved using an RF switch (EVAL-ADRF5044, Analog Devices), limiting the optical frequency switching time to a nominal 50 ns\cite{rfswitchdata}. EOMs were used to phase modulate at 1560 nm (MPZ-LN-10, iXblue) for frequency generation and at 780 nm (NIR-MPX800, iXblue) in the atomic-lock sub-system.  Optical spectra were measured using a scanning Fabry-Perot interferometer (SA200-5B, Thorlabs). Optical amplitude modulation was achieved using an AOM (T-M110-0.2C2J-3-F2P, Gooch and Housego) which was measured, at 1560 nm, to have a rise and fall time of 16.7 $\pm$ 0.2 and 18.5 $\pm$ 0.2 ns respectively. Because the efficiency of sum-frequency generation is proportional to the square of the electric field\cite{nonlinear}, to achieve 10 to 90 \% amplitude modulation at 780 nm requires only a \char`~32 to \char`~79 \% modulation at 1560 nm. Thus, at 780 nm, the rise and fall time was measured to be 14.2 $\pm$ 0.3 ns and 16.4 $\pm$ 0.2 ns. A biased InGaAs detector (DET10N2, Thorlabs) and 4 GHz oscilloscope (DSO9404A, Agilent) was used for measurements at both 1560 and 780 nm. The amplitude extinction ratio of the AOM was measured to be 60 $\pm$ 1 dB at 1560 nm, thus, due to the nonlinear efficiency, the extinction ratio at 780 nm is calculated to be 120 $\pm$ 2 dB. Indeed, at 780 nm, the extinction ratio exceeded the 80 dB dynamic range of the power meter (S132C, Thorlabs). A temperature-controlled, fiber-coupled periodically-poled lithium niobate waveguide (WH-0780-000-F-BC, NTT Electronics), was used to convert from 1560 to 780 nm. A 1 W erbium-doped fiber amplifier (EDFA) (ML1-EYFA-CW-SLM-P-OEM-SOA-1560, NKT), 2 W EDFA (F-CEFA-759-00, NKT),  and 3 W EDFA (YEDFA-PM-EM-3W-FC/APC-FC/APC-0, Orion Laser) were used to amplify the light to 500 mW and 2 W before the first FBG and the AOMs respectively. Maximal optical power was input to the FBG to allow minimal RF power (\char`~ -7 dBm) to be applied to the EOM, thus minimizing the amplitude of the EOM harmonics. These were not filtered by the FBG as the FBG reflection bandwidth (50 GHz) was too high and the EOM frequency range used (7 to 15 GHz) was too low. RF synthesizer evaluation boards (LMX2594EVM and LMX2595EVM, Texas Instruments), direct-digital synthesizer evaluation boards (EVAL-AD9914, Analog Devices), an arbitrary waveform generator (SDG6022X, Siglent) and a signal generator (E82667D, Keysight) were used to generate the required RF frequencies. The evaluation boards were programmed using microprocessors (PIC18F47K42 and PIC18F27K42, Microchip) and triggering was achieved using a CompactRIO Single-Board Controller (sbRio 9627, National Instruments). The experimental pulse sequence used to measure Rabi oscillations of the Raman transition (Fig. 5a) is shown in Figure 5c.
\begin{figure}
\centering
\includegraphics[width=0.95\linewidth]{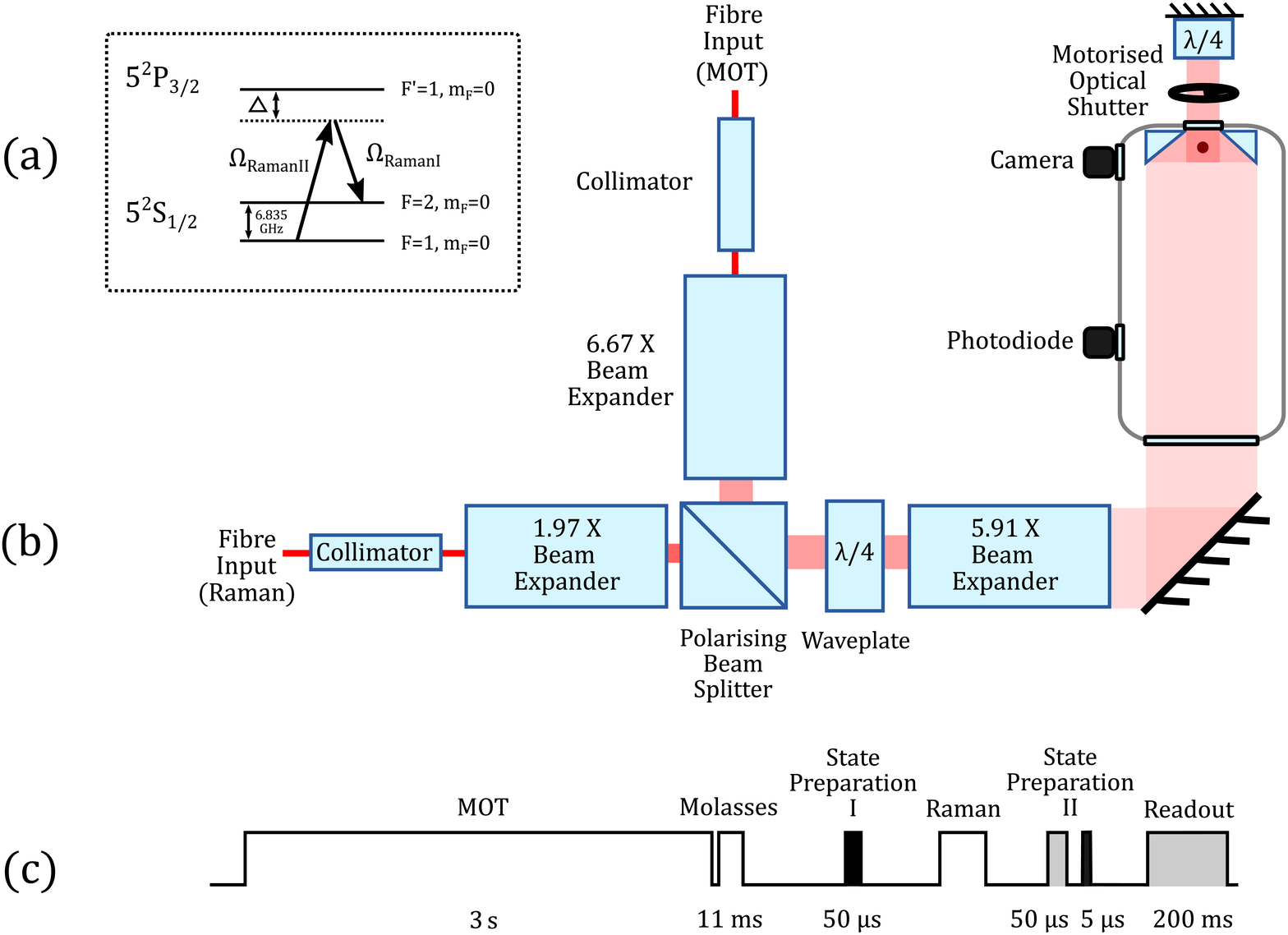}
\caption{(a) Raman transition energy level diagram where $\Delta$  is the Raman detuning, (b) configuration of vacuum chamber, sensors and free-space optics, and (c) experimental pulse sequence (not to scale). Black and grey pulses are resonant with F=1 to F'=2 and F=2 to F'=3 transitions of the \textsuperscript{87}Rb D\textsubscript{2} line respectively.}
\label{fig:freespace}
\end{figure}
The laser system was connected to the free-space section (Fig. 5b) via optical fibers. Here, the light was collimated, expanded and a quarter wave plate used to convert from linear to circular polarization. The light was coupled into a vacuum chamber and a magneto-optical trap (MOT) beam geometry was formed using light reflected from four prisms and a mirror. A quarter wave plate inserted before the mirror ensured retro-reflected beams maintained the same circular polarization handedness and a motorized optical shutter (SHB1T, Thorlabs) enabled blocking of the mirror and quarter-wave plate during Raman and readout pulses. The Raman and MOT beams had approximate beam diameters of 17 and 58 mm and powers of 341 and 539 mW respectively. Magnetic fields were generated by passing current through wire coils surrounding the chamber. 

Atomic state readout was achieved by measuring fluorescence with an unbiased photodiode (SM05PD1A, Thorlabs) and data was recorded with an oscilloscope (4262, Pico Technology). The atomic cloud was dropped \char`~15 cm before readout, and the Raman pulses occurred during free-fall.

A fast photodiode (125G-010HR-FC, Osi Optoelectronics) was used to measure the 6.835 GHz beat frequency of the laser during Raman pulse output. For comparison, the same frequency was also generated by mixing the RF frequencies input to the EOM. The phase noise of both signals was measured with a spectrum analyzer (N9030B, Keysight).
\section{Results}
\label{sec:Results}
Figure 6 shows the optical spectrum of the laser when outputting Raman pulses. Multiple spurious frequencies were observed due to sum frequency generation of the undesired carrier and EOM sideband components. 
\begin{figure} [H]
\centering
\includegraphics[width=0.7\linewidth]{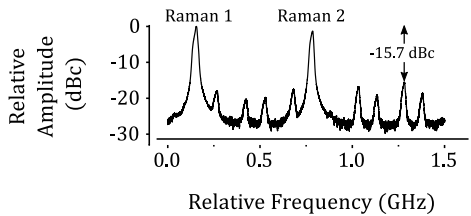}
\caption{Measurement of laser optical spectrum during Raman pulses using a scanning Fabry-Perot interferometer with a 1.5 GHz free spectral range. A Raman detuning of $\Delta$ = 500 MHz was used, and the data was averaged four times.}
\label{fig:raman}
\end{figure}
Overall, undesired optical frequencies were suppressed by \char`~15.7 dB at 780 nm. However, at the time of measurement, one of the FBGs had one of four thermoelectric coolers non-functional, likely reducing its performance. Suppression was \char`~4.3 dB below that achieved by IQ modulation\cite{IQ}, however, unlike with IQ modulators, performance can be further improved by adding additional FBGs. This is the first single-seed laser system providing both cooling and Raman light that avoids generating the Raman frequencies using a double-sideband method. Therefore, the suppression of undesired frequencies during Raman pulse sequences is improved by $>$ 12 dB compared with previous single-seed systems\cite{compact1560,compact780,onboard}.

Figure 7 shows the single-sideband phase noise of the optical and RF beat notes. As can be seen, the phase noise of the optical signal closely follows the RF source up to a \char`~20 kHz offset without the use of phase-locking electronics. This demonstrates successful RF to optical conversion.
\begin{figure} [H]
\centering
\includegraphics[width=0.55\linewidth]{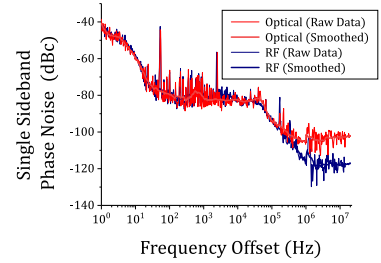}
\caption{Measurement of single-sideband phase noise of the optical and RF beat notes at 6.835 GHz using a spectrum analyzer.}
\label{fig:phase}
\end{figure}
Figure 8 shows Rabi oscillations of a copropagating two-photon Raman transition in ensembles of cold rubidium atoms obtained by varying the Raman pulse length. As can be seen, only when both Raman I and Raman II optical frequencies are applied, are oscillations observed. This demonstrates the successful cooling, trapping, manipulation and detection of atoms with the single-seed laser system, exhibiting its applicability to atom interferometry and quantum sensing. The contrast of the signal is limited by residual magnetic fields, beam alignment, spontaneous emission, the atomic cloud temperature and the finite size of the cloud within the Gaussian laser beam profile, none of which were fully optimized for this demonstration.
\begin{figure} [H]
\centering
\includegraphics[width=0.55\linewidth]{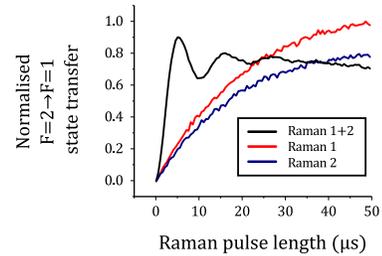}
\caption{Rabi oscillations of copropagating two-photon Raman transition obtained by varying the Raman pulse length. A Raman detuning of $\Delta$ = 500 MHz was used, and the data was averaged four times.}
\label{fig:rabi}
\end{figure}
\section{Conclusion}
\label{sec:Conclusion}
A wideband, agile, optical frequency generation scheme based on fiber Bragg grating technology was successfully demonstrated through the realization of a robust, alignment-free, single-seed, fibre laser system. The system was used to successfully laser cool and measure two-photon Raman transitions in ensembles of rubidium atoms. It exhibited exceptional modulation capabilities: the output frequency could be single-sideband modulated over a \char`~40 GHz range with a 50 ns frequency switching time; fast frequency and amplitude sweeping was enabled using DDS technology; and it had a 14.2 $\pm$ 0.3 ns amplitude rise time, 16.4 $\pm$ 0.2 ns fall time with an extinction ratio of 120 $\pm$ 2 dB. The output power was \char`~0.5 W per output channel, limited by the output power of the EDFA and power handling capabilities of the AOM.  An output power of \char`~1 W is expected in future iterations by upgrading to 5 W EDFAs and AOMs.

Compared with previous single-seed laser systems for rubidium atom interferometry\cite{compact1560}, the single-sideband modulation range was increased by a factor of 40 and the frequency switching time by a factor of 4000. This both increases the measurement bandwidth and can broaden the range of atom interferometry techniques available with single-seed lasers. 

Future versions could also enable filtering of the EOM harmonics by reducing the FBG reflection bandwidth and increasing the EOM frequency range such that the harmonics fall outside the reflection bandwidth. Multiple EOMs and FBGs could also be used with the same seed laser to generate multiple independent outputs, enabling, for example, the generation of separate Raman I and II beams. 

The system could be easily miniaturized to enable portable applications of atom interferometry and, in principle, the same agile frequency generation scheme could also be applied to other atomic species or quantum systems using similar components at different wavelengths. In future, this could help facilitate a new generation of portable quantum devices.
\section{Funding}
This work was funded by EPSRC under EP/T001046/1 as part of the UK National Quantum Technologies Programme.
\section{Acknowledgments}
We thank Farzad Hayati, Aisha Kaushik, Yu-Hung Lien and Ben Stray for many useful discussions during the development of the laser system.

\section{Disclosures}
\medskip

\noindent\textbf{Disclosures.} The authors declare no conflicts of interest.

\bibliography{BraggLaserArxiv}

\begin{thebibliography}{10}
\newcommand{\enquote}[1]{``#1''}

\bibitem{atomint}
M.~Kasevich and S.~Chu, \enquote{Atomic interferometry using stimulated raman
  transitions,} {\protect\JournalTitle{Phys. Rev. Lett.}} \textbf{67}, 181
  (1991).

\bibitem{rotation}
T.~L. Gustavson, A.~Landragin, and M.~A. Kasevich, \enquote{Rotation sensing
  with a dual atom-interferometer sagnac gyroscope,}
  {\protect\JournalTitle{Class. Quantum Grav.}} \textbf{17}, 2385 (2000).

\bibitem{gravity}
M.~Kasevich and S.~Chu, \enquote{Measurement of the gravitational acceleration
  of an atom with a light-pulse atom interferometer,}
  {\protect\JournalTitle{Appl. Phys. B}} \textbf{17}, 321--332 (1992).

\bibitem{gradio}
J.~M. McGuirk, G.~T. Foster, J.~B. Fixler, M.~J. Snadden, and M.~A. Kasevich,
  \enquote{Sensitive absolute-gravity gradiometry using atom interferometry,}
  {\protect\JournalTitle{Phys. Rev. A}} \textbf{65}, 033608 (2002).

\bibitem{equiv}
P.~Asenbaum, C.~Overstreet, M.~Kim, J.~Curti, and M.~A. Kasevich,
  \enquote{Atom-interferometric test of the equivalence principle at the
  10$-$12 level,} {\protect\JournalTitle{Phys. Rev. Lett.}} \textbf{125},
  191101 (2020).

\bibitem{mobilegrav}
C.~Freier, M.~Hauth, V.~Schkolnik, B.~Leykauf, M.~Schilling, H.~Wziontek, H.-G.
  Scherneck, J.~Müller, and A.~Peters, \enquote{Mobile quantum gravity sensor
  with unprecedented stability,} {\protect\JournalTitle{J. Phys.: Conf. Ser.}}
  \textbf{723}, 012050 (2016).

\bibitem{civil}
A.~Hinton, M.~Perea-Ortiz, J.~Winch, J.~Briggs, S.~Freer, D.~Moustoukas,
  S.~Powell-Gill, C.~Squire, A.~Lamb, C.~Rammeloo, B.~Stray, G.~Voulazeris,
  L.~Zhu, A.~Kaushik, Y.-H. Lien, A.~Niggebaum, A.~Rodgers, A.~Stabrawa,
  D.~Boddice, S.~R. Plant, G.~W. Tuckwell, K.~Bongs, N.~Metje, and M.~Holynski,
  \enquote{A portable magneto-optical trap with prospects for atom
  interferometry in civil engineering,} {\protect\JournalTitle{Phil. Trans. R.
  Soc. A}} \textbf{375}, 20160238 (2017).

\bibitem{space2}
K.~Bongs, M.~Holynski, and Y.~Singh, \enquote{$\phi$ in the sky,}
  {\protect\JournalTitle{Nature Phys}} \textbf{11}, 615--617 (2015).

\bibitem{chirp}
B.~Cheng, P.~Gillot, S.~Merlet, and F.~P.~D. Santos, \enquote{Influence of
  chirping the raman lasers in an atom gravimeter: Phase shifts due to the
  raman light shift and to the finite speed of light,}
  {\protect\JournalTitle{Phys. Rev. A}} \textbf{92}, 063617 (2015).

\bibitem{onboard}
F.~Theron, O.~Carraz, G.~Renon, N.~Zahzam, Y.~Bidel, M.~Cadoret, and
  A.~Bresson, \enquote{Narrow linewidth single laser source system for onboard
  atom interferometry,} {\protect\JournalTitle{Appl. Phys. B}} \textbf{118},
  1--5 (2015).

\bibitem{compact1560}
B.~Battelier, B.~Barrett, L.~Fouché, L.~Chichet, L.~Antoni-Micollier,
  H.~Porte, F.~Napolitano, J.~Lautier, A.~Landragin, and P.~Bouyer,
  \enquote{Development of compact cold-atom sensors for inertial navigation,}
  \url{https://arxiv.org/abs/1605.02454}.

\bibitem{compact780}
J.~Fang, J.~Hu, X.~Chen, H.~Zhu, L.~Zhou, J.~Zhong, J.~Wang, and M.~Zhan,
  \enquote{Narrow linewidth single laser source system for onboard atom
  interferometry,} {\protect\JournalTitle{Opt. Express}} \textbf{26},
  1586--1596 (2018).

\bibitem{serrodyne}
D.~M.~S. Johnson, J.~M. Hogan, S.~w.~Chiow, and M.~A. Kasevich,
  \enquote{Broadband optical serrodyne frequency shifting,}
  {\protect\JournalTitle{Opt. Lett.}} \textbf{35}, 745--747 (2010).

\bibitem{extrafreq}
O.~Carraz, R.~Charrière, M.~Cadoret, N.~Zahzam, Y.~Bidel, and A.~Bresson,
  \enquote{Phase shift in an atom interferometer induced by the additional
  laser lines of a raman laser generated by modulation,}
  {\protect\JournalTitle{Phys. Rev. A}} \textbf{86}, 033605 (2012).

\bibitem{IQ}
L.~Zhu, Y.~Lien, A.~Hinton, A.~Niggebaum, C.~Rammeloo, K.~Bongs, and
  M.~Holynski, \enquote{Application of optical single-sideband laser in raman
  atom interferometry,} {\protect\JournalTitle{Opt. Express}} \textbf{26},
  6542--6553 (2018).

\bibitem{IQbias}
X.~Li, L.~Deng, X.~Chen, M.~Cheng, S.~Fu, M.~Tang, and D.~Liu,
  \enquote{Modulation-format-free and automatic bias control for optical iq
  modulators based on dither-correlation detection,}
  {\protect\JournalTitle{Opt. Express}} \textbf{25}, 9333--9345 (2017).

\bibitem{nonlinear}
R.~Boyd (Academic Press, 2008).

\bibitem{braggreview}
S.~Dewra, V.~Plaha, and A.~Grover, \enquote{Fabrication and applications of
  fiber bragg grating- a review,} {\protect\JournalTitle{Adv. Eng. Tec. Appl.}}
  \textbf{4}, 15--25 (2015).

\bibitem{fbgdata}
{TeraXion Inc.}, \enquote{{TFN Ultra Narrowband Tunable Optical Filter
  Datasheet},}
  \url{https://teraxion.blob.core.windows.net/media/1355/teraxion-tfn-specsheet.pdf}.

\bibitem{eomdata}
iXblue, \enquote{{MPX and MPZ series Datasheet},}
  \url{https://photonics.ixblue.com/sites/default/files/2020-02/MPX\%2BMPZ\%20SERIES.pdf}.

\bibitem{MTS}
D.~J. McCarron, S.~A. King, and S.~L. Cornish, \enquote{Modulation transfer
  spectroscopy in atomic rubidium,} {\protect\JournalTitle{Meas. Sci.
  Technol.}} \textbf{19}, 105601 (2018).

\bibitem{rfswitchdata}
{Analog Devices Inc.}, \enquote{{ADRF5044 Datasheet},}
  \url{https://www.analog.com/media/en/technical-documentation/data-sheets/adrf5044.pdf}.

\end{thebibliography}

\end{document}